\documentclass[12pt,a4paper]{article}
\usepackage[T1]{fontenc}
\usepackage{times}

\usepackage{amsmath}
\usepackage{amsthm}
\usepackage{amssymb}
\usepackage[utf8]{inputenc}
\usepackage{cite}
\def\jgpapersize{}

\def\Real{{\mathbb R}}

\def\innerprod(#1,#2){{\left<#1\,,\,#2\right>}}
\def\Set#1{{\left\{#1\right\}}}

\def\quadtext#1{\quad\textup{#1}\quad}
\def\quadand{\quadtext{and}}
\def\pfrac#1#2{\frac{\partial #1}{\partial #2}}

\def\dfrac#1#2{\frac{d #1}{d #2}}
\def\Dfrac#1#2{\frac{D #1}{d #2}}

\def\sumtripleindex#1#2#3{
\begin{array}[t]{c}
\mbox{\scriptsize$#3$}\\
{\displaystyle\sum}\\
\mbox{\scriptsize$#1$}\\[-0.3em]
\mbox{\scriptsize$#2$}
\end{array}}

\def\sumtripleindex#1#2#3{
\sum^{#3}_{\substack{{#1}\\{#2}}}
\end{array}}

\usepackage{fancyhdr}
\fancypagestyle{firststyle}
{
   \fancyhf{}
   
\fancyfoot[C]{1}
\fancyfoot[R]{\raisebox{10em}{\footnotesize\bf File: \jobname.tex}}
}
\allowdisplaybreaks

\usepackage{color}

\def\DEcomma{{\,,}}
\def\DEFull{{\,.}}
\def\DENone{{\,,}}

\def\Cdot{{\dot{C}}}
\def\Cddot{{\ddot{C}}}

\def\taumin{\tau^-}
\def\taumax{\tau^+}

\def\iMa{{\cRed{a}}}
\def\iMb{{\cRed{b}}}
\def\iMc{{\cRed{c}}}
\def\iMd{{\cRed{d}}}
\def\iMe{{\cRed{e}}}
\def\iMf{{\cRed{f}}}
\def\iSa{{\cRed{i}}}

\def\iPa{{\cBlue{A}}}
\def\iPb{{\cBlue{B}}}

\def\iMa{{{\mu}}}
\def\iMb{{{\nu}}}
\def\iMc{{{\rho}}}
\def\iMd{{{\sigma}}}
\def\iMe{{{\alpha}}}
\def\iMf{{{\beta}}}
\def\iSa{{{i}}}

\def\iPa{{{A}}}
\def\iPb{{{B}}}

\def\Vr{{\boldsymbol r}}
\def\Asum{{\overline \Afield}{}}
\def\Fsum{{\overline F}{}}
\def\Afield{{\cal A}}

\def\tot{{\textup{tot}}}

\def\MLSI/{{ML--SI}}

\def\VE{{\boldsymbol E}}
\def\VB{{\boldsymbol B}}

\def\part{{\textup{part}}}
\def\LagR{{\cal L}}

\def\affiliation{}
\def\homepage{\thanks}
\begin{document}
\jgpapersize

\title{
Maxwell-Lorentz without self interactions: Conservation of energy and momentum.}

\author{Jonathan Gratus$^{1,2,3}$} 

\maketitle

\affiliation{$^1$ 
  Department of Physics,
  Lancaster University,
  Lancaster LA1 4YB,
  United Kingdom,
}

\affiliation{$^2$ 
The Cockcroft Institute,
Sci-Tech Daresbury,
Daresbury WA4 4AD,
United Kingdom,
}

\homepage{$^3$ https://orcid.org/0000-0003-1597-6084}

\thispagestyle{firststyle} 

\begin{abstract}
Since a classical charged point particle radiates energy and momentum
it is argued that there must be a radiation reaction force. Here
we present an action for the Maxwell-Lorentz without self interactions
model, where each particle only responds to the fields of the other
charged particles. The corresponding
stress-energy tensor automatically conserves energy and momentum in
Minkowski and other appropriate spacetimes.
\end{abstract}

\section{Introduction}

Over the last century there has been significant scientific work debating how
elementary charged particles respond to their own electromagnetic
field, and the corresponding question of the electromagnetic mass
\cite{Rohrlich65,jackson1999classical,PhysRevD.99.096001,
frisch2004inconsistency,lyle2010self,ferris2011origin}.  
The standard conclusion is that a charged particle
obeys the Abraham–Lorentz–Dirac force. This has well known
pathologies. There exist runaway solutions where particles accelerate
forever without a force \cite{poisson1999introduction}. 
The run-away solutions can be avoided
by using the critical submanifold \cite{Spohn_2000}, but this leads to
pre-acceleration, where a particle moves before a force is applied. 
Alternative approaches
include using a delay equation \cite{PhysRevD.99.096001}, the
Eliezer–Ford–O’Connell equation \cite{garcia2015mathematical}, considering the
Landau-Lipsitz equation as fundamental (and not an approximation)
\cite{rohrlich2001correct}, 
and replacing Maxwell's
equation with Born-Infield \cite{burton2011born} 
or Bopp-Podolski \cite{gratus2015self,lazar2020second,zayats2014self}.

For practical purposes it has often been simpler to ignore any
radiation reaction and assume that each particle responds via the
Lorentz force to the external electromagnetic field and the fields of
all the other charged particles. After all in most cases the sum of
the fields of the other particles will dominate any radiation
reaction.  For example Klimontovich \cite[Page
  42]{davidson2001physics} used such a model to make statistical
predictions about the nature of plasmas.  However with the new intense
laser fields available soon, for example with ELI, it is projected
that the radiation reaction will not only be detectable, but may dominant the
motion of electrons in plasma
\cite{burton2014aspects,kravets2013radiation,di2012extremely}.

The model where each particle does not respond to its own field 
is referred to as Maxwell-Lorentz without self-interaction
(\MLSI/) \cite{lazarovici2018against,bauer2013maxwell}.  
The \MLSI/ model is usually
criticised on the grounds that it does not account for the damping
force experienced by an accelerated charge. Underlying this criticism
is the observation that an accelerating charge radiates and that
the energy of this radiation must come from somewhere. Usually it is
assumed to the kinetic energy of the charged particle. Likewise
the radiation has a momentum and this requires a back-force on the
radiating particle.

In this letter it is demonstrated that the \MLSI/ model does in-fact conserve the total energy and momentum. As a result, the criticisms above are no longer justified. Whatever energy is required to accelerate the charged particles must be balanced by the energy in the electromagnetic field which is lost. This model make specific predictions. It predicts that no radiation reaction will be observed even in the above laser-plasma interactions where it would otherwise be detectable.

\vspace{1em}

In this model, the universe consists of a finite number of point
charges. Each charge's 4--acceleration is determined by the sum of the
electromagnetic fields of all the other particles. The approach is
covariant so it applies to any spacetime. There is no
``external'' electromagnetic field. 
The model is described by an action, or equivalently a Lagrangian. As
well as deriving the dynamic equations for the particles and fields,
the Lagrangian can also be used to derive the stress-energy
tensor. There are two ways of deriving the stress-energy tensor from a
Lagrangian. By varying the Lagrangian with respect to the
metric one obtains the ``Hilbert'' stress-energy tensor. However one
can use the Noether fields and the Belenfante-Rosen modification
to derive the ``Belenfante'' stress-energy tensor.

Since all the fields in this action are dynamic, i.e. there are no
external fields, other than gravity, there are three important
consequences \cite{gratus2012conservation}: The Hilbert and
Belenfante-Rosen definitions of the stress-energy tensor are equal,
the stress-energy tensor is symmetric and it is covariantly
conserved. Thus all Killing symmetries of the spacetime will give rise
to conserved currents. This establishes the claim that for Minkowski
spacetime, energy, momentum and angular momentum are conserved.

\vspace{1em}

The letter is organised as follows: We start by giving 
the Lagrangian of the \MLSI/ model, the corresponding dynamic equations and
the stress-energy tensors. The derivation of these results is placed in
the appendix. 
We then give some predictions of this model.
In the next section we list additional consequences of the model. 
We then conclude.


\section{The self-interaction free model}

\label{ch_Lag}

Let $M$ be spacetime with background metric $g_{\iMa\iMb}$, $\iMa,\iMb=0,1,2,3$
with signature $(-,+,+,+)$.  Consider a collection of $N$ particles 
$\Set{P_\iPa,\,\iPa=1,\ldots,N}$,
with masses $m_\iPa$ and charges $q_\iPa$. The
particles travel along worldlines $C_\iPa^{\iMa}(\tau)$ with proper
time \footnote{In cases where the particles do not accelerate to
  lightlike infinity, $\taumin_\iPa=-\infty$ and $\taumax_\iPa=\infty$. However we
  do not want to exclude the case when a particle accelerates to
  infinity in a finite proper time.} 
parameterisation $\tau$ with $\taumin_\iPa<\tau<\taumax_\iPa$ so
that \footnote{
We use the summation convention on the spacetime index $\iMa=0,1,2,3$, no
summation on the particle index $\iPa$. 
The particle index $\iPa$ is placed either
high or low for convenience.
}
the 4-velocity $\Cdot_\iPa^{\iMa}(\tau)=\dfrac{}{\tau} C_\iPa^\iMa$ 
satisfies $\Cdot_\iPa^{\iMa}\,\Cdot^\iPa_{\iMa}=-1$.
One needs to consider $N$ electromagnetic fields $F^\iPa_{\iMa\iMb}$ which are
generated by $N$ potentials $\Afield^\iPa_{\iMa}$ so that 
$F^\iPa_{\iMa\iMb}=\partial_{\iMa} \Afield^\iPa_{\iMb} - \partial_{\iMb} \Afield^\iPa_{\iMa}$. 

The action depends on the $N$ worldlines $C_\iPa^{\iMa}(\tau)$, the $N$ 
fields $\Afield^\iPa_{\iMa}$ and the metric $g_{\iMa\iMb}$ is given formally by
\begin{equation}
\begin{aligned}
\lefteqn{
S[C_\iPa^{\iMa},\Afield^\iPa_{\iMa},g_{\iMa\iMb}]
}
\qquad&
\\&=
\sum_{\iPa=1}^N
\frac{m_\iPa}{2}\, \int_{\taumin_\iPa}^{\taumax_\iPa}\!
\Cdot_{\iPa}^{\iMa}(\tau) \, \Cdot^\iPa_{\iMa}(\tau)\ 
d\tau
\\&\quad
-
\sum_{\iPa=1}^N
\sum_{\iPb,\iPb\ne\iPa}
{q_\iPa}\, \int_{\taumin_\iPa}^{\taumax_\iPa}
\Cdot_{\iPa}^{\iMa}(\tau)\, \Afield^\iPb_{\iMa}|_{C_\iPa(\tau)}\ 
d\tau
\\&\quad
-
\frac14 
\sum_{\iPa=1}^N
\sum_{\iPb,\iPb\ne\iPa}
\int_M
F^\iPa_{\iMa\iMb} \, F_\iPb^{\iMa\iMb} \ 
\omega\ d^4 x 
\DEcomma
\end{aligned}
\label{Lag_Lag}
\end{equation}
where $\omega=\sqrt{(-\det{g})}$ and 
$\displaystyle\sum_{\iPb,\iPb\ne\iPa}$ refers to the double sum
over all $\iPb$ excluding $\iPa$,
i.e. $\displaystyle\sum_{\iPb,\iPb\ne\iPa}=\sum_{\iPb=1}^{\iPa-1}+\sum_{\iPb=\iPa+1}^N$. We
state that (\ref{Lag_Lag}) is the formal action as any actual integral
would likely diverge. To get a convergent integral it is necessary to
integrate over a compact region $U\subset M$ so that 
\begin{equation}
\begin{aligned}
\lefteqn{S[C_\iPa^{\iMa},\Afield^\iPa_{\iMa},g_{\iMa\iMb},U]}
\qquad&
\\&=
\sum_{\iPa=1}^N
\frac{m_\iPa}{2}\, \int_{I_\iPa(U)}\!
\Cdot_{\iPa}^{\iMa}(\tau) \, \Cdot^\iPa_{\iMa}(\tau)\ 
d\tau
\\&\quad
-
\sum_{\iPa=1}^N
\sum_{\iPb,\iPb\ne\iPa}
{q_\iPa}\, \int_{I_\iPa(U)}
\Cdot_{\iPa}^{\iMa}(\tau)\, \Afield^\iPb_{\iMa}|_{C_\iPa(\tau)}\ 
d\tau
\\&\quad
-
\frac14 
\sum_{\iPa=1}^N
\sum_{\iPb,\iPb\ne\iPa}
\int_U
F^\iPa_{\iMa\iMb} \, F_\iPb^{\iMa\iMb} \ 
\omega\ d^4 x 
\DEcomma
\end{aligned}
\label{Lag_Lag_U}
\end{equation}
where $I_\iPa(U)=\Set{\tau|C_\iPa(\tau)\in U}$. An alternative method is
to use a test function, which we do below in (\ref{Lag_finite}).

We can write the action $S$ in terms of a distributional Lagrangian
$L$
\begin{align}
S = \int_U L \,\omega \, d^4x
\label{Lag_Lag_Int}
\DEcomma
\end{align}
where
\begin{equation}
\begin{aligned}
L 
&= 
\sum_{\iPa=1}^N
\frac{m_\iPa}{2}\,\omega^{-1}\, \int_{\taumin_\iPa}^{\taumax_\iPa}\!
\Cdot_{\iPa}^{\iMa} \, \Cdot^\iPa_{\iMa}\,
\delta\big(x-C_{\iPa}(\tau)\big)\,
d\tau
\\&\quad
-
\sum_{\iPa=1}^N
\sum_{\iPb,\iPb\ne\iPa}
{q_\iPa} \,\omega^{-1}\,\int_{\taumin_\iPa}^{\taumax_\iPa}
\Cdot_{\iPa}^{\iMa}\, \Afield^\iPb_{\iMa}
\,
\delta\big(x-C_{\iPa}(\tau)\big)\,
d\tau
\\&\quad-
\frac14 
\sum_{\iPa=1}^N
\sum_{\iPb,\iPb\ne\iPa}
F^\iPa_{\iMa\iMb}\, F_\iPb^{\iMa\iMb}
\DEFull
\end{aligned}
\label{Lag_Lag_val}
\end{equation}

Varying $S$ with respect to $C_\iPa^{\iMa}$ gives the Lorentz force equation
for the particle $\iPa$,
\begin{align}
m_\iPa \,\Dfrac{\Cdot_{\iPa}^{\iMa} }{\tau} 
=
q_\iPa \,\Cdot^\iPa_{\iMb}\sum_{\substack{\iPb,\iPb\ne\iPa}} F_\iPb^{\iMb\iMa}
\DEcomma
\label{Lag_Lorentz_Force}
\end{align}
where 
$\Dfrac{}{\tau} \Cdot_{\iPa}^{\iMa}=\Cddot_{\iPa}^{\iMa} + 
\Gamma^\iMa_{\iMb\iMc} \Cdot_{\iPa}^{\iMb}\Cdot_{\iPa}^{\iMc}$.

Varying $S$ with respect to $\Afield^\iPa_{\iMa}$ gives 
\begin{align}
\sum_{\iPb,\iPb\ne\iPa} \!\nabla_{\iMb} F^{\iMb\iMa}_\iPb 
= 
\sum_{\iPb,\iPb\ne\iPa} q_\iPb\,\omega^{-1} \int_{\taumin_\iPb}^{\taumax_\iPb}
\Cdot^{\iMa}_\iPb\, \delta(x-C_{\iPb}(\tau))\, d\tau
\DEFull
\label{Lag_Max_alt}
\end{align}
This is equivalent to the electromagnetic field $F^\iPb_{\iMa\iMb}$ being generated
by the particle $P_\iPb$.
\begin{align}
\nabla_{\iMb} F^{\iMb\iMa}_\iPb 
= 
q_\iPb\,\omega^{-1} \int_{\taumin_\iPb}^{\taumax_\iPb}
\Cdot^{\iMa}_\iPb\, \delta(x-C_{\iPb}(\tau))\, d\tau
\DEFull
\label{Lag_Max}
\end{align}
Differentiating the Lagrangian density $L\,\omega$ 
with respect to the metric 
\begin{align}
T^{\iMa\iMb} 
=
2\,\omega^{-1} \,\pfrac{(L\,\omega)}{g_{\iMa\iMb}}
=
2\,\pfrac{L}{g_{\iMa\iMb}} 
+ 
2\,L\, \omega^{-1} \,\pfrac{\omega}{g_{\iMa\iMb}}
\DENone
\label{Lag_T_Hil_def}
\end{align}
gives the ``Hilbert'' stress-energy tensor
\begin{equation}
\begin{aligned}
T^{\iMa\iMb} 
&=
\omega^{-1}
\sum_{\iPa=1}^N \int_{\taumin_\iPa}^{\taumax_\iPa}
m_\iPa \, \delta^{(4)}\big(x-C_{\iPa}(\tau)\big)
\, \Cdot^{\iMa}_\iPa\, \Cdot^{\iMb}_{\iPa} \,  d\tau
\\&\qquad
+
\sum_{\iPa=1}^N 
\sum_{\iPb,\iPb\ne\iPa}
\Big( 
F^{\iMa}_{\iPa}{}_{\iMc} F_{\iPb}^{\iMb\iMc} 
-
\tfrac14 g^{\iMa\iMb} F_{\iPa}^{\iMc\iMd} F^{\iPb}_{\iMc\iMd}\Big)
\DEFull
\end{aligned}
\label{Lag_SEM_alt}
\end{equation}
This is the same as the ``Belenfante-Rosen'' stress-energy tensor 
\begin{equation}
\begin{aligned}
T^{\iMa}{}_{\iMb}
&=
L\,\delta^{\iMa}_{\iMb}
-
2\,\sum_{\iPa=1}^N F^\iPa_{\iMb\iMc}\,\pfrac{L}{F^\iPa_{\iMa\iMc}}
\\&\quad
- 
\sum_{\iPa=1}^N \Afield^\iPa_{\iMb}\,\pfrac{L}{\Afield^\iPa_{\iMa}}
+
\sum_{\iPa=1}^N \Cdot^{\iMa}_\iPa\,\pfrac{L}{\Cdot^{\iMb}_{\iPa}}
\\&\quad
-
\sum_{\iPa=1}^N \delta^{\iMa}_{\iMb}\,\delta\big(x-C_{\iPa}(\tau)\big)
\pfrac{L}{\big(\delta(x-C_{\iPa}(\tau))\big)}
\DEcomma
\end{aligned}
\label{Lag_BR}
\end{equation}
where we define
\begin{equation}
\begin{aligned}
\delta\big(x-C_{\iPa}(\tau)\big)
&\pfrac{}{\big(\delta(x-C_{\iPa}(\tau))\big)}
\int \LagR\, \delta(x-C_{\iPa}(\tau))\, d\tau
\\&=
\int \LagR\, \delta(x-C_{\iPa}(\tau))\, d\tau
\DEFull
\end{aligned}
\label{Lag_def_d_delta}
\end{equation}
This last term of (\ref{Lag_BR}) may look contrived. However it arises
from diffeomorphism invariance of the action. The
details are outlined in the appendix.

The stress-energy tensor is a total stress-energy tensor, therefore it
has the symmetry of the
indices
\begin{align}
T^{\iMa\iMb} = T^{\iMb\iMa}
\DENone
\label{Lag_T_sym}
\end{align}
and the divergenceless condition, which is also known as being
covariantly conserved
\begin{align}
\nabla_{\iMa} T^{\iMa\iMb} = 0
\DEFull
\label{Lag_Div_T}
\end{align}
The derivations of (\ref{Lag_Lorentz_Force})-(\ref{Lag_Div_T}) are given
in the appendix. 

\vspace{1em}

We observe that both the Lagrangian $L$ and the stress-energy tensor
$T^{\iMa\iMb}$ are well defined distributions. That is for any
test function $\varphi$ and test tensor $\varphi_{\iMa\iMb}$, the integrals
\begin{align}
\int_M L\,\varphi\,d^4x
\quadand
\int_M T^{\iMa\iMb}\,\varphi_{\iMa\iMb}\,d^4x
\quadtext{are finite.}
\label{Lag_finite}
\end{align}
This assumes that no two worldlines ever intersect. 
To see this observe that, away from all the particles, there are
only the electromagnetic fields $F^{\iPa}_{\iMa\iMb}$, which are all
finite. Thus if $\varphi$ and $\varphi_{ab}$ have support which does
not intersect any worldline, the integrals in (\ref{Lag_finite}) are finite.

Along a particle $P_\iPa$, the line integrals are clearly bounded.
Choose an adapted coordinate system $(x^0,\ldots,x^3)$
such that $C^0_\iPa(\tau)=\tau$ and $C^\iSa_\iPa(\tau)=0$ for $\iSa=1,2,3$.
Let $r\in\Real$ be the spatial distance given by
$r=((x^1)^2+(x^2)^2+(x^3)^2)^{1/2}$. Then approximately
$F^{\iPa}_{\iMa\iMb}(r)\approx q_\iPa\,r^{-2}$. That is it does not go to
infinity faster than $r^{-2}$. However in polar coordinates
$\omega=r^2\sin\theta$ so $F^{\iPa}_{\iMa\iMb}(r)\omega$ is
bounded. Also $F^{\iPb}_{\iMa\iMb}(r)$ for $P_\iPb\ne P_\iPa$ are all
bounded so the integrals in (\ref{Lag_finite}) are all bounded.

This contrasts with the standard Lagrangian for electromagnetism which
contains $F_{\iMa\iMb}\,F^{\iMa\iMb}$ for a single field. In this case
$F_{\iMa\iMb}\,F^{\iMa\iMb}\approx r^{-4}$ which is not defined as a
distribution.

\section{Predictions of the \MLSI/ model}
\label{ch_Predict}

The principle prediction of this model is that no radiation reaction
observed. As stated in the introduction, forthcoming laser-plasma
experiments will have electromagnetic fields of sufficient intensities
that radiation reaction, it is exist, should be detectable. Thus if
charged particle motion is consistent with the Lorentz force and not
consistent with the models which include radiation reaction, then this
would provide strong evidence for \MLSI/.

Another interesting prediction is for the case when the universe has a
non-trivial topology. In Minkowski,
Friedmann–Lemaître–Robertson–Walker or many other spacetimes it is
impossible for a charged particle $P_{\iPa}$ to interact with its own
electromagnetic radiation after it has been produced. By contrast if
the universe is a 3--torus times time, then $P_{\iPa}$ will see
multiple copies of itself and will respond many times as its radiation
goes round the torus. This model predicts that whereas $P_{\iPa}$ will
respond multiple times to the radiation of the other particles
$P_\iPb\ne P_\iPa$, it will never respond to its own radiation.  The
same observation could, in principle, be made in our universe by looking at how
charged particles behave in the ergosphere of a rotating black
hole. In this case part of the radiation $F^\iPa_{\iMa\iMb}$ will
rotate faster than $P_\iPa$ and hence have the opportunity to interact
again with $P_\iPa$. However, in the \MLSI/ model it will not do so.

This contrasts with the case when a charged particle $P_{\iPa}$ sees itself in a
perfect electrical conductor. A perfect electrical conductor
is, itself, an approximation of a real metal made of electrons and
protons. These will respond to the initial charged particle and in
turn construct a field which can be seen by $P_{\iPa}$.

This model also predicts that, in a universe with just a single charged 
particle, there would be no electromagnetic field produced and the
particle will simply undergo geodesic motion. In Minkowski spacetime,
the value of the single homogeneous field
$\Afield^\iMa_{\textup{hom},\iPa}$, where $\iPa=1$, is irrelevant as it does
not affect the motion of the particle $P_\iPa$.

\section{Other Consequences of the \MLSI/ model}

{\bf The Cauchy problem:} 
The \MLSI/ is also a well defined Cauchy problem. We can find Cauchy
surfaces on which we can prescribe the initial positions $C_\iPa^{\iMa}$ and
velocities $\Cdot_\iPa^{\iMa}$ as well as the electromagnetic field
$F^\iPa_{\iMa\iMb}$. The standard theories about the Cauchy problem
for Maxwell's equations and the Lorentz force equation, 
then state we can find the subsequent values.

\vspace{1em}

{\bf A type of regularisation:} 
We may think of \MLSI/ as a regularisation of the Lagrangian. Let
$F^\tot_{\iMa\iMb}=\sum_{\iPa=1}^N F^{\iPa}_{\iMa\iMb}$ then for
points away from any particles we may replace (\ref{Lag_Lag_val}) and
(\ref{Lag_SEM_alt}) with
\begin{align}
L 
=
\frac14 
F^\tot_{\iMa\iMb}\, F_\tot^{\iMa\iMb}
-
\tfrac14
\sum_{\iPa=1}^N
F^\iPa_{\iMa\iMb}\, F_\iPa^{\iMa\iMb}
\DENone
\label{Conc_Lag_reg}
\end{align}
and
\begin{equation}
\begin{aligned}
T^{\iMa\iMb} 
&=
\Big(\tfrac12 F^{\iMa}_{\tot}{}_{\iMc} F_{\tot}^{\iMc\iMb} + 
\tfrac18 g^{\iMa\iMb} F_{\tot}^{\iMc\iMd} F^{\tot}_{\iMc\iMd}
\Big)
\\&\quad
-
\sum_{\iPa=1}^N 
\Big( 
\tfrac12 F^{\iMa}_{\iPa}{}_{\iMc} F_{\iPa}^{\iMc\iMb} + 
\tfrac18 g^{\iMa\iMb} F_{\iPa}^{\iMc\iMd} F^{\iPa}_{\iMc\iMd}\Big)
\DEFull
\end{aligned}
\label{Conc_Lag_SEM_reg}
\end{equation}
However (\ref{Conc_Lag_reg}) and (\ref{Conc_Lag_SEM_reg})
do not extend to the worldlines as each term on the right hand sides
diverge as $\approx r^{-4}$ as one approaches the worldline. Thus
they are not individually distributions.

\vspace{1em}

{\bf Liénard-Wiechart fields:} 
In Minkowski spacetime we can solve Maxwell's equations
(\ref{Lag_Max}) using the
Liénard-Wiechart potentials.
\def\tauR{{\tau_{\textup{R}}}}
\begin{align}
\Afield^\iMa_\iPa (x)
&=
\Afield^\iMa_{\textup{hom},\iPa} + 
\frac{q_\iPa\,\Cdot^{\iMa}_\iPa(\tauR)}
{\Cdot^{\iMb}_\iPa(\tauR) \big(x-C^\iPa_{\iMb}(\tauR)\big)}
\DEcomma
\label{Lag_LW_Field}
\end{align}
where $\tauR$ is the retarded time and $\Afield^\iMa_{\textup{hom},\iPa}$ is
a solution to the source free Maxwell equation 
$\nabla_{\iMb} F^{\iMa\iMb}_{\textup{hom},\iPa}=0$ with
$F^{\textup{hom},\iPa}_{\iMa\iMb}
=\partial_{\iMa} \Afield^{\textup{hom},\iPa}_{\iMb} 
- \partial_{\iMb} \Afield^{\textup{hom},\iPa}_{\iMa}$. 

\vspace{1em}

{\bf Violation of weak energy condition:}
Finally, we observe that the \MLSI/ stress-energy tensor does not satisfy the 
weak energy condition. In Minkowski spacetimes, let the particle pass
$P_\iPa$ through the origin, $C_\iPa^{\iMa}(\tau_0)=(0,0,0,0)$ with
$\Cdot_\iPa^{\iMa}(\tau_0)=(1,0,0,0)$. 
Consider an observer at the point $x^\iMa=(r,x^1,x^2,x^3)$ where 
$r=((x^1)^2+(x^2)^2+(x^3)^2)^{1/2}$ is small.
In this case
$T_{\iMa\iMb} \Cdot^{\iMa}_\iPa\Cdot^{\iMb}_\iPa = T^{00}=
\VE_\iPa\cdot\overline\VE_\iPa + \VB_\iPa\cdot\overline\VB_\iPa$ where 
$\VE_\iPa$ and $\VB_\iPa$ are the electric and magnetic fields due to
particle $P_\iPa$ and
$\overline\VE_\iPa=\sum_{\iPb,\iPb\ne\iPa}\VE_\iPb$ 
and $\overline\VB_\iPa=\sum_{\iPb,\iPb\ne\iPa}\VB_\iPb$ are
the fields due to all the other particles. Now $T^{00}$ is dominated
by the Coulomb term, 
$T^{00}\approx -q_\iPa\,r^{-2}\,\hat{\Vr}\cdot\overline\VE_\iPa$. Now
choosing $\Vr=\pm \VE_\iPa$, depending on the sign of $q_\iPa$, then
$T^{00}<0$. The violation of the weak energy condition does not cause
problems for the classical theory, but may cause issues for the
corresponding quantum theory. This violation is also true for other
classical theories such as Bopp-Podolski \cite{cuzinatto2018bopp}.

\section{Discussion and Conclusion}
\label{ch_Conc}

In this letter we have presented the case why Maxwell-Lorentz without
self interaction is the best model for the dynamics of charged
particles. It has many advantages. 

The action (\ref{Lag_Lag}) is a total action, that is all the fields
are dynamic. As a result all the equations of motion are derivable
from the action and there are no requirements for constitutive relations
or equations of state. Another advantage is that the stress-energy
tensor derived from varying the Lagrangian with respect to the metric
is equivalent to that derived from the Noether current with the
Belenfante-Rosen procedure. This tensor is symmetric in its indices
and divergenceless and thus gives rise to conserved quantities whenever
a Killing vector is present.

The \MLSI/ makes predictions about the behaviour of electrons in
plasmas which may be testable soon.

The Lagrangian is also useful in that it only requires a finite
quantity of information to specify, namely the masses and charges of
the particles. Although, in general this quantity is large. 
This contrasts with for example, the cold plasma model
of charge, which requires a infinite quantity of information.

This model is certainly simpler than the different models which
include radiation reaction. Some of these can also be derived from an
action, but which may involve doubling the phase space \cite{barone2007lagrangian}.

The main disadvantage is the violation of the weak
stress-energy tensor, which although not relevant for the classical
theory, poses challenges for quantisation.

Another criticisms of the \MLSI/ model is the $N$--fold increase in
electromagnetic fields \cite{lazarovici2018against}. Thus to set up
the Cauchy problem requires $N$ initial conditions for the $N$
particles as well as $N$ initial conditions for the $N$ fields. In
Minkowski spacetimes, this can be seen by the $N$ homogeneous fields
$\Afield^\iMa_{\textup{hom},\iPa}$ given in (\ref{Lag_LW_Field}). However if
we could argue that one can set the homogeneous fields
$\Afield^\iMa_{\textup{hom},\iPa}=0$, then we can replace Maxwell's equation
(\ref{Lag_Max_alt}) with the Liénard-Wiechart fields. Thus we are left
with only differential difference equations.  Thus the \MLSI/ approach
is similar to the Feynmann-Wheeler theory
\cite{RevModPhys.17.157,bauer2013existence}, in that particles only
communicate with other particles. However since, in the Minkowski
case, only the retarded potential is used, the system is manifestly
casual.

Even the results of future experiments indicate the existence of
radiation reaction, the \MLSI/ model is a useful tool.
It may be extended for the interaction of point dipoles and
quadrupoles 
\cite{ellis1966electromagnetic, dixon1970dynamics, gratus2018correct, 
gratus2020distributional}. Here the multipoles only
respond to the electromagnetic fields of the other particles. 
This is particularly important because the question of
the self-interaction of multipoles is very challenging.
Although these multipole models are not derived from
a Lagrangian, that they can be described by a function
distributional stress-energy tensor is very useful.

\section{Acknowledgements}
The author acknowledges support provided
by STFC (Cockcroft Institute, ST/G008248/1 and ST/P002056/1), 
and 
 EPSRC (Lab in a Bubble, EP/N028694/1).
He would also like to thank Prof. Robin Tucker, Dr. Spyridon Talaganis
and Alex Warwick for their
helpful advice.

\bibliographystyle{unsrt}
\bibliography{NoSelf}

\newpage


\appendix

\section{Proofs of the formulae}
\label{ch_Pfs}
We use the standard results
\begin{align}
\pfrac{\omega}{g_{\iMa\iMb}} 
=
-\tfrac12 \omega^{-1} \pfrac{(\det{g})}{g_{\iMa\iMb}} 
=
-\tfrac12 (\det{g}) \omega^{-1} {g^{\iMa\iMb}} 
=
\tfrac12 \omega^{} {g^{\iMa\iMb}} 
\label{Pfs_d_omega_gab}
\end{align}
so
\begin{equation}
\begin{aligned}
\partial_\iMa\,\omega
&=
\partial_\iMa\,\sqrt{(-\det{g})}
=
- \tfrac12\omega^{-1} \partial_\iMa\,(\det{g})
=
- \tfrac12\omega^{-1} (\det{g})\,g^{\iMb\iMc}\,\partial_\iMa g_{\iMb\iMc} 
=
\tfrac12\omega\,g^{\iMb\iMc}\,\partial_\iMa g_{\iMb\iMc} 
\\&=
\tfrac12\omega g^{\iMd\iMb}\,
(\partial_\iMa g_{\iMb\iMd} + \partial_\iMb g_{\iMa\iMd} 
- \partial_\iMd g_{\iMa\iMb})
=
\delta^{\iMb}_{\iMc}\,\tfrac12\omega g^{\iMd\iMc}\,
(\partial_\iMa g_{\iMb\iMd} + \partial_\iMb g_{\iMa\iMd} 
- \partial_\iMd g_{\iMa\iMb})
=
\omega\,\delta^{\iMb}_{\iMc}\Gamma^\iMc_{\iMb\iMa}
=
\omega\,\Gamma^\iMb_{\iMb\iMa}
\end{aligned}
\label{Pfs_d_a_omega}
\end{equation}
and
\begin{align}
\pfrac{g^{\iMc\iMd}}{g_{\iMa\iMb}}
=
- g^{\iMc\iMe} \pfrac{g_{\iMe\iMf}}{g_{\iMa\iMb}} g^{\iMf\iMd}
=
- g^{\iMc\iMe} (\delta^{\iMa}_{\iMe}\delta^{\iMb}_{\iMf}) g^{\iMf\iMd}
=
- g^{\iMc\iMa}  g^{\iMb\iMd}
\DEFull
\label{Pfs_d_gcd_gab}
\end{align}

For the following, let $\Asum^\iPa_\iMa=\sum_{\iPb,\iPb\ne \iPa} \Afield^\iPb_\iMa$, 
$\Fsum^\iPa_{\iMa\iMb} = \partial_\iMa\Asum^\iPa_\iMa -
\partial_\iMb\Asum^\iPa_\iMa$ then (\ref{Lag_Lag}) becomes
\begin{align}
S
&=
\sum_{\iPa=1}^N
\int_{\taumin_\iPa}^{\taumax_\iPa}
\Big(
\frac{m_\iPa}{2}\, 
\Cdot_{\iPa}^{\iMa} \, \Cdot^\iPa_{\iMa}\ 
-
{q_\iPa}\,
\Cdot_{\iPa}^{\iMa}\, 
\Asum^\iPa_{\iMa}|_{C_{\iPa}(\tau)}\ 
\Big)
d\tau
-
\frac14 
\sum_{\iPa=1}^N
\int_M
F^\iPa_{\iMa\iMb} \, \Fsum_\iPa^{\iMa\iMb} \ 
\omega\ d^4 x 
\DEFull
\label{Lag_Lag_bar}
\end{align}


\begin{proof}[Proof that varying \eqref{Lag_Lag} with respect to
    $C_{\iPa}$ gives \eqref{Lag_Lorentz_Force}]

Let $L_\iPa=\tfrac12{m_\iPa}\, \Cdot_{\iPa}^{\iMa} \,
\Cdot^\iPa_{\iMa}\ 
-{q_\iPa}\,\Cdot_{\iPa}^{\iMa}\, \Asum^\iPa_{\iMa}|_{C_{\iPa}(\tau)}$.
From the standard Euler-Lagrange formula, from (\ref{Lag_Lag_bar}) we have
\begin{align*}
0
&=
\pfrac{L_\iPa}{C^\iMa_\iPa} - 
\dfrac{}{\tau}\Big(\pfrac{L_\iPa}{\Cdot^\iMa_\iPa}\Big)
=
\tfrac12\,m_\iPa\,\Cdot^{\iMb}_\iPa\,\Cdot^{\iMc}_{\iPa}\, \partial_\iMa g_{\iMb\iMc}
-
{q_\iPa}\,\Cdot^{\iMb}_{\iPa}\,\partial_\iMa\Asum^\iPa_\iMb
-
\dfrac{}{\tau}\Big(m_\iPa\,\Cdot^{\iMb}_{\iPa}\,g_{\iMa\iMb} 
- 
{q_\iPa} \,\Asum^\iPa_\iMa\Big)
\\&=
\tfrac12\,m_\iPa\,\Cdot^{\iMb}_\iPa\,\Cdot^{\iMc}_{\iPa}\, \partial_\iMa g_{\iMb\iMc}
-
{q_\iPa}\,\Cdot^{\iMb}_{\iPa}\,\partial_\iMa\Asum^\iPa_\iMb
-
m_\iPa\,\Cddot^{\iMb}_{\iPa}\,g_{\iMa\iMb} 
- 
m_\iPa\,\Cdot^{\iMb}_{\iPa}\Cdot^{\iMc}\,\partial_\iMc g_{\iMa\iMb}
+
{q_\iPa} \,\Cdot^{\iMc}\,\partial_\iMc \Asum^\iPa_\iMa
\\&=
-
{q_\iPa}\,\Cdot^{\iMb}_{\iPa}\,\Fsum^\iPa_{\iMa\iMb}
-
m_\iPa\,\Cddot^{\iMb}_{\iPa}\,g_{\iMa\iMb} 
+ 
\tfrac12\,m_\iPa\,\Cdot^{\iMb}_\iPa\,\Cdot^{\iMc}_{\iPa}\, 
\big(
\partial_\iMc g_{\iMa\iMb}
+
\partial_\iMb g_{\iMa\iMc}
-
\partial_\iMa g_{\iMb\iMc}
\big)
\\&=
-
{q_\iPa}\,\Cdot^{\iMb}_{\iPa}\,\Fsum^\iPa_{\iMa\iMb}
-
m_\iPa\,\big(\Cddot^{\iMb}_{\iPa}\,g_{\iMa\iMb} 
+ 
\Cdot^{\iMb}_\iPa\,\Cdot^{\iMc}_{\iPa}\,\Gamma^{\iMd}{}_{\iMb\iMc} g_{\iMa\iMd} 
\big)
\\&=
{q_\iPa}\,\Cdot^{\iMb}_{\iPa}\,\Fsum^\iPa_{\iMb\iMa}
-
m_\iPa\,\big(\Cddot^{\iMb}_{\iPa}\,g_{\iMa\iMb} 
+ 
\Cdot^{\iMd}_\iPa\,\Cdot^{\iMc}_{\iPa}\,\Gamma^{\iMb}{}_{\iMd\iMc} g_{\iMa\iMb} 
\big)
\\&=
{q_\iPa}\,\Cdot^{\iMb}_{\iPa}\,\Fsum^\iPa_{\iMb\iMa}
-
m_\iPa\,g_{\iMa\iMb}\,\Cdot^{\iMc}_{\iPa}\nabla_{\iMc}\Cdot^{\iMb}_{\iPa}
\end{align*}
hence (\ref{Lag_Lorentz_Force}) follows.
\end{proof}


\begin{proof}[Proof that varying \eqref{Lag_Lag} with respect to
    $\Afield^{\iPa}_{\iMa}$ gives \eqref{Lag_Max_alt}]
Observe from (\ref{Pfs_d_a_omega})
\begin{align*}
\partial_\iMb (F^{\iMb\iMa}_{\iPa}\, \omega)
&=
(\partial_\iMb F^{\iMb\iMa}_{\iPa})\, \omega
+
F^{\iMb\iMa}_{\iPa}\,\partial_\iMb \omega
=
(\partial_\iMb F^{\iMb\iMa}_{\iPa})\, \omega
+
F^{\iMb\iMa}_{\iPa}\,\Gamma^\iMc_{\iMc\iMb}\, \omega
=
\omega\,\big(\partial_\iMb F^{\iMb\iMa}_{\iPa}
+
F^{\iMc\iMa}_{\iPa}\,\Gamma^\iMb_{\iMb\iMc}
+
F^{\iMb\iMc}_{\iPa}\,\Gamma^\iMa_{\iMb\iMc}
\big)
\\&=
\omega\,\nabla_\iMb F^{\iMb\iMa}_{\iPa}
\DEFull
\end{align*}

From (\ref{Lag_Lag_val}) we have
\begin{align*}
L\omega 
&= 
\sum_{\iPa=1}^N
\frac{m_\iPa}{2}\, \int_{\taumin_\iPa}^{\taumax_\iPa}\!
\Cdot_{\iPa}^{\iMa} \, \Cdot^\iPa_{\iMa}\,
\delta\big(x-C_{\iPa}\big)\,
d\tau
-
\sum_{\iPa=1}^N
\sum_{\iPb,\iPb\ne\iPa}
{q_\iPb} \int_{\taumin_\iPb}^{\taumax_\iPa}
\Cdot_{\iPb}^{\iMa}\, \Afield^\iPa_{\iMa} 
\,
\delta\big(x-C_{\iPb}\big)\,
d\tau
\\&\qquad
-
\frac12
\sum_{\iPa=1}^N
\sum_{\iPb,\iPb\ne\iPa}
\partial_\iMa \Afield^\iPa_{\iMb}\, F_\iPb^{\iMa\iMb} \, 
\omega
\DEcomma
\end{align*}
hence
\begin{align*}
0 
&=
\pfrac{(L\omega)}{\Afield^{\iPa}_\iMa}
-
\partial_\iMb \pfrac{(L\omega)}{(\partial_\iMb \Afield^{\iPa}_\iMa)}
=
-
\sum_{\iPb,\iPb\ne\iPa}
{q_\iPb} \int_{\taumin_\iPa}^{\taumax_\iPa}
\Cdot_{\iPb}^{\iMa}\, \delta(x-C_{\iPb})
\,d\tau
+
\partial_\iMb 
\Big(\sum_{\iPb,\iPb\ne\iPa}
F_\iPb^{\iMb\iMa}\omega
\Big)
\\&=
\sum_{\iPb,\iPb\ne\iPa}
\Big(
-
{q_\iPb} \int_{\taumin_\iPa}^{\taumax_\iPa}
\Cdot_{\iPb}^{\iMa}\, \delta(x-C_{\iPb})
\,d\tau
+
\partial_\iMb 
\big(
F_\iPb^{\iMb\iMa}\omega
\big)
\Big)
\\&=
\sum_{\iPb,\iPb\ne\iPa}
\Big(
-
{q_\iPb} \int_{\taumin_\iPa}^{\taumax_\iPa}
\Cdot_{\iPb}^{\iMa}\, \delta(x-C_{\iPb})
\,d\tau
+
\omega
\,\nabla_\iMb 
F_\iPb^{\iMb\iMa}
\Big)
\DEFull
\end{align*}
\end{proof}

\begin{proof}[Proof that \eqref{Lag_Max_alt} implies \eqref{Lag_Max}]
Sum (\ref{Lag_Max_alt}) over all $\iPa$ (and dividing by the number of
particles) gives
\begin{align*}
0
&=
\frac1N \sum_{\iPa=1}^N
\sum_{\iPb,\iPb\ne\iPa} 
\Big(
\nabla_{\iMb} F^{\iMb\iMa}_\iPb 
-
\omega^{-1}
q_\iPb \int_{\taumin_\iPb}^{\taumax_\iPb}
\Cdot^{\iMa}_\iPa\, \delta(x-C_{\iPa}(\tau))\, d\tau
\Big)
\\&=
\sum_{\iPa=1}^N
\Big(
\nabla_{\iMb} F^{\iMb\iMa}_\iPa 
-
\omega^{-1}
q_\iPa \int_{\taumin_\iPa}^{\taumax_\iPb}
\Cdot^{\iMa}_\iPa\, \delta(x-C_{\iPa}(\tau))\, d\tau
\Big)
\DEFull
\end{align*}
Subtracting (\ref{Lag_Max_alt}) gives (\ref{Lag_Max}).
\end{proof}


\begin{proof}[Proof that \eqref{Lag_T_Hil_def} gives \eqref{Lag_SEM_alt}]
Writing (\ref{Lag_Lag_bar}) with $g_{\iMa\iMb}$ explicit we have
\begin{equation}
\begin{aligned}
\omega\, L
&= 
\sum_{\iPa=1}^N
\frac{m_\iPa}{2}\, \int_{\taumin_\iPa}^{\taumax_\iPa}\!
\Cdot_{\iPa}^{\iMa} \, \Cdot_\iPa^{\iMb}\,g_{\iMa\iMb}\,
\delta\big(x-C_{\iPa}(\tau)\big)\,
d\tau
-
\sum_{\iPa=1}^N
{q_\iPa} \,\int_{\taumin_\iPa}^{\taumax_\iPa}
\Cdot_{\iPa}^{\iMa}\, \Asum^\iPa_{\iMa}
\,
\delta\big(x-C_{\iPa}(\tau)\big)\,
d\tau
\\&\qquad-
\frac14 
\sum_{\iPa=1}^N
F^\iPa_{\iMa\iMb}\, \Fsum^\iPa_{\iMc\iMd}\, g^{\iMa\iMc}\,g^{\iMb\iMd}\,\omega
\DEFull
\end{aligned}
\label{Pfs_omega_L}
\end{equation}
When differentiating with respect to $g_{\iMa\iMb}$ we ignore the
constraint $g_{\iMa\iMb}=g_{\iMb\iMa}$. I.e. we assume that
$g_{\iMa\iMb}$ is independent of $g_{\iMb\iMa}$, when $\iMa\ne\iMb$. Thus
\begin{align*}
\pfrac{}{g_{\iMa\iMb}} 
(\Cdot_{\iPa}^{\iMc} \, \Cdot_\iPa^{\iMd}\,g_{\iMc\iMd})
&=
\Cdot_{\iPa}^{\iMc} \, \Cdot_\iPa^{\iMd}\,\delta^{\iMa}_\iMc\,\delta^{\iMb}_{\iMd}
=
\Cdot_{\iPa}^{\iMa} \, \Cdot_\iPa^{\iMb}
\DEFull
\end{align*}
Hence differentiating the first term of (\ref{Pfs_omega_L}) gives
the first term of (\ref{Lag_SEM_alt}). The second term of
(\ref{Pfs_omega_L}) in independent of $g_{\iMa\iMb}$ so does not
contribute to $T^{\iMa\iMb}$. 

For the last term
\begin{align*}
\pfrac{}{g_{\iMa\iMb}} (F^\iPa_{\iMc\iMd}\,\Fsum^\iPa_{\iMe\iMf}\, g^{\iMc\iMe}\, g^{\iMd\iMf})
&=
-
F^\iPa_{\iMc\iMd}\,\Fsum^\iPa_{\iMe\iMf}\, 
(g^{\iMc\iMa}\, g^{\iMb\iMe}\, g^{\iMd\iMf}
+g^{\iMc\iMe}\, g^{\iMd\iMa}\, g^{\iMb\iMf})
\\&=
-
(F^\iPa_{\iMc\iMd}\,\Fsum_\iPa^{\iMb\iMd} \,g^{\iMc\iMa}
+F^\iPa_{\iMc\iMd}\,\Fsum_\iPa^{\iMc\iMb}\, g^{\iMd\iMa}
)
=
-
(F^{\iMa}_{\iPa\iMd}\,\Fsum_\iPa^{\iMb\iMd}
+F_{\iPa\iMc}^{\iMa}\,\Fsum_\iPa^{\iMc\iMb} )
=
-2
F^{\iMa}_{\iPa\iMc}\,F^{\iMb\iMc}
\DEFull
\end{align*}
Hence the last term of (\ref{Pfs_omega_L}) gives
\begin{align*}
-\tfrac12 
\pfrac{}{g_{\iMa\iMb}} \Big(\sum_{\iPa=1}^N 
F^\iPa_{\iMc\iMd}\,\Fsum_\iPa^{\iMc\iMd}\,\omega\Big)
=
\omega\sum_{\iPa=1}^N
\Big(
F^{\iMa}_{\iPa\iMc}\,\Fsum^{\iMc\iMb}
-
\tfrac12
F^\iPa_{\iMc\iMd}\,\Fsum_\iPa^{\iMc\iMd}
\Big)
\DEFull
\end{align*}
Hence (\ref{Lag_SEM_alt}).

\end{proof}


\begin{proof}[Proof that \eqref{Lag_BR} implies \eqref{Lag_SEM_alt}.]
Taking each term in (\ref{Lag_BR}) in turn. Second term:
\begin{align*}
-2\sum_{\iPa=1}^N F^\iPa_{\iMb\iMc}\,\pfrac{L}{F^\iPa_{\iMa\iMc}}
&=
\sum_{\iPa=1}^N F^\iPa_{\iMb\iMc}
\,{\Fsum_\iPa^{\iMa\iMc}}
\DEFull
\end{align*}
Third term:
\begin{align*}
-\sum_{\iPa=1}^N \Afield^\iPa_{\iMb}\,\pfrac{L}{\Afield^\iPa_{\iMa}}
&=
\sum_{\iPa=1}^N
\Afield^\iPa_{\iMb}\,
\sum_{\iPb,\iPb\ne\iPa}
{q_\iPa} \,\omega^{-1}\,\int_{\taumin_\iPa}^{\taumax_\iPa}
\Cdot_{\iPb}^{\iMa}\,
\,
\delta\big(x-C_{\iPa}(\tau)\big)\,
d\tau
\\&=
\sum_{\iPa=1}^N
{q_\iPa} \,\omega^{-1}\,\int_{\taumin_\iPa}^{\taumax_\iPa}
\Cdot_{\iPa}^{\iMa}\,
\Asum^\iPa_{\iMb}\,
\,
\delta\big(x-C_{\iPa}(\tau)\big)\,
d\tau
\DEFull
\end{align*}
Fourth term:
\begin{align*}
\sum_{\iPa=1}^N
\Cdot^{\iMa}_\iPa\,\pfrac{L}{\Cdot^{\iMb}_{\iPa}}
=
\sum_{\iPa=1}^N
m_\iPa\,\omega^{-1}\, \int_{\taumin_\iPa}^{\taumax_\iPa}\!
\Cdot_{\iPa}^{\iMa} \, \Cdot^\iPa_{\iMb}\,
\delta\big(x-C_{\iPa}(\tau)\big)\,
d\tau
-
\sum_{\iPa=1}^N
{q_\iPa} \,\omega^{-1}\,\int_{\taumin_\iPa}^{\taumax_\iPa}
\Cdot_{\iPa}^{\iMa}\,
\Asum^\iPa_{\iMb}\,
\,
\delta\big(x-C_{\iPa}(\tau)\big)\,
d\tau
\DEFull
\end{align*}
Fifth term:
\begin{align*}
-\sum_{\iPa=1}^N &\delta^{\iMa}_{\iMb}\,\delta\big(x-C_{\iPa}(\tau)\big)
\pfrac{L}{\Big(\delta\big(x-C_{\iPa}(\tau)\big)\Big)}
\\&=
-\delta^{\iMa}_{\iMb}\,
\sum_{\iPa=1}^N
\frac{m_\iPa\,\omega^{-1}}{2}\, \int_{\taumin_\iPa}^{\taumax_\iPa}\!
\Cdot_{\iPa}^{\iMa} \, \Cdot^\iPa_{\iMa}\,
\delta\big(x-C_{\iPa}(\tau)\big)\,
d\tau
+
\sum_{\iPa=1}^N
{q_\iPa} \,\omega^{-1}\,\int_{\taumin_\iPa}^{\taumax_\iPa}
\Cdot_{\iPa}^{\iMa}\, \Asum^\iPa_{\iMa}
\,
\delta\big(x-C_{\iPa}(\tau)\big)\,
d\tau
\DEFull
\end{align*}
Adding these together gives (\ref{Lag_SEM_alt}).
\end{proof}


\begin{proof}[Proof of \eqref{Lag_Div_T}]
\def\Talt{{\cal T}}

Given test functions $\phi_\iMb$ then for any tensor
$\Talt^{\iMa\iMb}$ then
\begin{align*}
\int_M (\nabla_\iMa \Talt^{\iMa\iMb}) \,\phi_\iMb\,\omega \, d^4x
&=
\int_M (\partial_\iMa \Talt^{\iMa\iMb}) \,\phi_\iMb\,\omega \, d^4x
+
\int_M \Talt^{\iMc\iMb}\,\Gamma^\iMa_{\iMa\iMc} \,\phi_\iMb\,\omega \, d^4x
+
\int_M \Talt^{\iMa\iMc}\,\Gamma^\iMb_{\iMa\iMc} \,\phi_\iMb\,\omega \, d^4x
\\&=
-
\int_M \Talt^{\iMa\iMb} \,\partial_\iMa (\phi_\iMb\,\omega) \, d^4x
+
\int_M \Talt^{\iMc\iMb}\,(\partial_\iMc\omega) \phi_\iMb \, d^4x
+
\int_M \Talt^{\iMa\iMc}\,\Gamma^\iMb_{\iMa\iMc} \,\phi_\iMb\,\omega \, d^4x
\\&=
-
\int_M \Talt^{\iMa\iMb} \,(\partial_\iMa\phi_\iMb)\,\omega \, d^4x
+
\int_M \Talt^{\iMa\iMc}\,\Gamma^\iMb_{\iMa\iMc} \,\phi_\iMb\,\omega \, d^4x
=
-
\int_M \Talt^{\iMa\iMb} \,(\nabla_\iMa\phi_\iMb)\,\omega \, d^4x
\DEFull
\end{align*}
For the first term of (\ref{Lag_SEM_alt}) we have
\begin{align*}
\int_M \nabla_\iMa &\bigg(\omega^{-1}\sum_{\iPa=1}^N \int_{\taumin_\iPa}^{\taumax_\iPa}
m_\iPa \, \delta^{(4)}\big(x-C_{\iPa}(\tau)\big)
\, \Cdot^{\iMa}_\iPa\, \Cdot^{\iMb}_{\iPa} \,  d\tau\bigg)
\,\phi_\iMb\,\omega \, d^4x
\\&=
-
\int_M 
\bigg(\sum_{\iPa=1}^N \int_{\taumin_\iPa}^{\taumax_\iPa}
m_\iPa \, \delta^{(4)}\big(x-C_{\iPa}(\tau)\big)
\, \Cdot^{\iMa}_\iPa\, \Cdot^{\iMb}_{\iPa} \,  d\tau\bigg)
\,(\nabla_\iMa\phi_\iMb)\, d^4x
\\&=
-
\sum_{\iPa=1}^N \int_{\taumin_\iPa}^{\taumax_\iPa}
m_\iPa 
\, \Cdot^{\iMa}_\iPa\, \Cdot^{\iMb}_{\iPa} 
\,(\nabla_\iMa\phi_\iMb)\, d\tau
=
-
\sum_{\iPa=1}^N \int_{\taumin_\iPa}^{\taumax_\iPa}
m_\iPa 
\, \Cdot^{\iMb}_{\iPa} 
\,\Dfrac{\phi_\iMb}{\tau}\, d\tau
\\&=
-
\sum_{\iPa=1}^N \int_{\taumin_\iPa}^{\taumax_\iPa}
m_\iPa 
\, \Cdot^{\iMb}_{\iPa} 
\Big(\dfrac{\phi_\iMb}{\tau}-\Gamma^\iMc_{\iMa\iMb}\Cdot^{\iMa}_{\iPa}\,\phi_\iMb\Big)\, d\tau
\\&=
\sum_{\iPa=1}^N \int_{\taumin_\iPa}^{\taumax_\iPa}
m_\iPa 
\,\Big(\Cddot^{\iMb}_{\iPa} 
+\Gamma^\iMc_{\iMa\iMb}\,\Cdot^{\iMb}_{\iPa}\,\Cdot^{\iMa}_{\iPa}\,
\Big)\phi_\iMb\, d\tau
\\&=
\sum_{\iPa=1}^N \int_{\taumin_\iPa}^{\taumax_\iPa}
m_\iPa 
\,\Dfrac{\Cdot^{\iMb}_{\iPa}}{\tau} 
\phi_\iMb\, d\tau
\DEFull
\end{align*}
Hence using the Lorentz force equation (\ref{Lag_Lorentz_Force})
\begin{align}
\nabla_\iMa \sum_{\iPa=1}^N \omega^{-1}\int_{\taumin_\iPa}^{\taumax_\iPa}
m_\iPb \, \delta^{(4)}\big(x-C_{\iPa}(\tau)\big)
\, \Cdot^{\iMa}_\iPa\, \Cdot^{\iMb}_{\iPa} \,  d\tau
&=
\sum_{\iPa=1}^N \omega^{-1}
\int_{\taumin_\iPa}^{\taumax_\iPa} m_\iPa 
\,\Dfrac{\Cdot^{\iMa}_{\iPa}}{\tau} \,\delta\big(x-C_{\iPa}(\tau)\big)
\,d\tau
\notag
\\&=
\sum_{\iPa=1}^N \omega^{-1}
\int_{\taumin_\iPa}^{\taumax_\iPa} 
q_\iPa \,\Cdot^\iPa_{\iMb} \,\Fsum_\iPa^{\iMb\iMa}
\delta\big(x-C_{\iPa}(\tau)\big)
\,d\tau
\DEFull
\label{Pfs_DelT_Lor}
\end{align}

For the second term of (\ref{Lag_SEM_alt}) we have
\begin{align*}
\nabla_\iMb
\Big( 
\Fsum^{\iMa}_{\iPa}{}_{\iMc} F_{\iPa}^{\iMb\iMc} 
-
\tfrac14 g^{\iMa\iMb} F_{\iPa}^{\iMc\iMd} \Fsum^{\iPa}_{\iMc\iMd}\Big)
&=
\big(\nabla_\iMb\Fsum^{\iMa}_{\iPa}{}_{\iMc}\big) F_{\iPa}^{\iMb\iMc} 
+ \Fsum^{\iMa}_{\iPa}{}_{\iMc} \big(\nabla_\iMb F_{\iPa}^{\iMb\iMc}\big) 
-\tfrac14 g^{\iMa\iMb} \big(\nabla_\iMb F_{\iPa}^{\iMc\iMd}\big) 
\Fsum^{\iPa}_{\iMc\iMd}
-\tfrac14 g^{\iMa\iMb} F_{\iPa}^{\iMc\iMd} \big(\nabla_\iMb \Fsum^{\iPa}_{\iMc\iMd}\big)
\\&=
\Fsum^{\iMa}_{\iPa}{}_{\iMc} \big(\nabla_\iMb F_{\iPa}^{\iMb\iMc}\big) 
+
\big(\nabla_\iMb\Fsum^{\iPa}_{\iMd\iMc}\big) g^{\iMa\iMd} F_{\iPa}^{\iMb\iMc} 
-\tfrac12 g^{\iMa\iMb} F_{\iPa}^{\iMc\iMd} \big(\nabla_\iMb \Fsum^{\iPa}_{\iMc\iMd}\big)
\\&=
\Fsum^{\iMa}_{\iPa}{}_{\iMc} \big(\nabla_\iMb F_{\iPa}^{\iMb\iMc}\big) 
+
\tfrac12 g^{\iMa\iMd} \Big(
\big(2\nabla_\iMb\Fsum^{\iPa}_{\iMd\iMc}\big) F_{\iPa}^{\iMb\iMc} 
-
F_{\iPa}^{\iMc\iMb} \big(\nabla_\iMd \Fsum^{\iPa}_{\iMc\iMb}\big)
\Big)
\\&=
\Fsum^{\iMa}_{\iPa}{}_{\iMc} \big(\nabla_\iMb F_{\iPa}^{\iMb\iMc}\big) 
+
\tfrac12 g^{\iMa\iMd} \Big(
\big(\nabla_\iMb\Fsum^{\iPa}_{\iMd\iMc}\big) F_{\iPa}^{\iMb\iMc} 
+
\big(\nabla_\iMc\Fsum^{\iPa}_{\iMd\iMb}\big) F_{\iPa}^{\iMc\iMb} 
+
F_{\iPa}^{\iMb\iMc} \big(\nabla_\iMd \Fsum^{\iPa}_{\iMc\iMb}\big)
\Big)
\\&=
\Fsum^{\iMa}_{\iPa}{}_{\iMc} \big(\nabla_\iMb F_{\iPa}^{\iMb\iMc}\big) 
+
\tfrac12 g^{\iMa\iMd} F_{\iPa}^{\iMb\iMc} \Big(
\nabla_\iMb\Fsum^{\iPa}_{\iMd\iMc} 
+
\nabla_\iMc\Fsum^{\iPa}_{\iMb\iMd}
+
\nabla_\iMd \Fsum^{\iPa}_{\iMc\iMb}
\Big)
\\&=
\Fsum^{\iMa}_{\iPa}{}_{\iMc} \big(\nabla_\iMb F_{\iPa}^{\iMb\iMc}\big) 
\DEFull
\end{align*}
Hence from Maxwell (\ref{Lag_Max}) 
\begin{align*}
\nabla_\iMb
\sum_{\iPa=1}^N \Big( 
\Fsum^{\iMa}_{\iPa}{}_{\iMc} F_{\iPa}^{\iMb\iMc} 
-
\tfrac14 g^{\iMa\iMb} F_{\iPa}^{\iMc\iMd} \Fsum^{\iPa}_{\iMc\iMd}\Big)
&=
\sum_{\iPa=1}^N \Fsum^{\iMa}_{\iPa}{}_{\iMc} \big(\nabla_\iMb F_{\iPa}^{\iMb\iMc}\big) 
=
q_\iPb\,\omega^{-1} \int_{\taumin_\iPb}^{\taumax_\iPb}
\Cdot^{\iMc}_\iPb\, \Fsum^{\iMa}_{\iPa}{}_{\iMc} \,\delta(x-C_{\iPb}(\tau))\, d\tau
\\&=
- \sum_{\iPa=1}^N q_\iPb\,\omega^{-1} \int_{\taumin_\iPb}^{\taumax_\iPb}
\Cdot_{\iMc}^\iPb\, \Fsum^{\iMc\iMa}_{\iPa} \,\delta(x-C_{\iPb}(\tau))\, d\tau
\DEFull
\end{align*}

Combining this with (\ref{Pfs_DelT_Lor}) gives (\ref{Lag_Div_T}).

\end{proof}


\subsection{Diffeomorphism invariance and the Noether formulation of
  stress-energy tensor}

The stress-energy can be derived in two ways from a Lagrangian. Either
from variation with respect to the metric (\ref{Lag_T_Hil_def})
or by using the Noether theorem, with the Belenfante-Rosen
procedure (\ref{Lag_BR}). That these two are the same
is due to: one, the action is diffeomorphism invariant
and two, there are no background fields, other than the
metric. These two conditions imply that the action is a total action,
i.e. all the fields it depends on are dynamic. Total actions also lead
to the dynamical equations, the index symmetry of the stress-energy tensor
(\ref{Lag_T_sym}) and that it is divergenceless (\ref{Lag_Div_T}). As
stated in the conclusion, however, such total actions are actually
quite rare.
The effect of background fields is detailed in \cite{gratus2012conservation}.

The details of (\ref{Lag_BR}) require discussion about diffeomorphisms
which is beyond the scope of this article. However (\ref{Lag_BR}) has
been demonstrated directly above. To see where the last term of
(\ref{Lag_BR}) originates 
from it is sufficient to observe that for a partial Lagrangian
\begin{align*}
S^\part = \int_{\taumin}^{\taumax} \LagR (\Cdot^\iMa,C^\iMa)\, d\tau
\end{align*}
and corresponding distributional Lagrangian
\begin{align*}
L^\part= \int_{\taumin}^{\taumax} \LagR (\Cdot^\iMa,C^\iMa)\,
\delta(x-C_{\iPb}(\tau))\,
d\tau
\DEcomma
\end{align*}
then the corresponding stress-energy tensor is given by the fourth
term of (\ref{Lag_BR}), i.e.
\begin{align*}
(T^\part){}^\iMa{}_{\iMb} = 
\int_{\taumin}^{\taumax} \Cdot^{\iMa}\,\pfrac{\LagR}{\Cdot^{\iMb}}
\delta(x-C_{\iPb}(\tau))\,
d\tau
\DEFull
\end{align*}
However, the first term of (\ref{Lag_BR}),
$L^\part\,\delta^{\iMa}_{\iMb}$, does not contribute to the
stress-energy tensor. This is removed due to (\ref{Lag_def_d_delta}),
which gives the following
\begin{align*}
L^\part
=
\delta\big(x-C(\tau)\big)
\pfrac{L^\part}{\big(\delta(x-C(\tau))\big)}
\DEFull
\end{align*}
Hence (\ref{Lag_BR}).


\end{document}